\documentclass[12pt]{article}
\usepackage{amsfonts}
\usepackage{amssymb}
\usepackage{graphics,psboxit,amsmath}
\usepackage{subfigure}
\usepackage{graphicx}
\usepackage{verbatim}
\usepackage{epic}

\def\hybrid{\topmargin 0pt \oddsidemargin 0pt 
        \headheight 0pt \headsep 0pt
        \textwidth 16,5cm 
        \textheight 23cm 
        \marginparwidth .875in
        \parskip 5pt plus 1pt \jot = 1.5ex}

\hybrid

\catcode`\@=11

\def\marginnote#1{}
\newcount\hour
\newcount\minute
\newtoks\amorpm
\hour=\time\divide\hour by60
\minute=\time{\multiply\hour by60 \global\advance\minute by-\hour}
\edef\standardtime{{\ifnum\hour<12 \global\amorpm={am}%
        \else\global\amorpm={pm}\advance\hour by-12 \fi
        \ifnum\hour=0 \hour=12 \fi
        \number\hour:\ifnum\minute<10 0\fi\number\minute\the\amorpm}}
\edef\militarytime{\number\hour:\ifnum\minute<10 0\fi\number\minute}

\def\draftlabel#1{{\@bsphack\if@filesw {\let\thepage\relax
   \xdef\@gtempa{\write\@auxout{\string
      \newlabel{#1}{{\@currentlabel}{\thepage}}}}}\@gtempa
   \if@nobreak \ifvmode\nobreak\fi\fi\fi\@esphack}
        \gdef\@eqnlabel{#1}}
\def\@eqnlabel{}
\def\@vacuum{}
\def\draftmarginnote#1{\marginpar{\raggedright\scriptsize\tt#1}}

\def\draft{\oddsidemargin -.5truein
        \def\@oddfoot{\sl preliminary draft \hfil
        \rm\thepage\hfil\sl\today\quad\militarytime}
        \let\@evenfoot\@oddfoot \overfullrule 3pt
        \let\label=\draftlabel
        \let\marginnote=\draftmarginnote
   \def\@eqnnum{(\theequation)\rlap{\kern\marginparsep\tt\@eqnlabel}%
\global\let\@eqnlabel\@vacuum} }

\def\draft2{
        \def\@oddfoot{\sl preliminary draft \hfil
        \rm\thepage\hfil\sl\today\quad\militarytime}
        \let\@evenfoot\@oddfoot \overfullrule 3pt
        \let\label=\draftlabel
        \let\marginnote=\draftmarginnote
   \def\@eqnnum{(\theequation)\rlap{\kern\marginparsep\tt\@eqnlabel}%
\global\let\@eqnlabel\@vacuum} }

\def\preprint{\twocolumn\sloppy\flushbottom\parindent 2em
        \leftmargini 2em\leftmarginv .5em\leftmarginvi .5em
        \oddsidemargin -.5in \evensidemargin -.5in
        \columnsep .4in \footheight 0pt
        \textwidth 10.in \topmargin -.4in
        \headheight 12pt \topskip .4in
        \textheight 6.9in \footskip 0pt
        \def\@oddhead{\thepage\hfil\addtocounter{page}{1}\thepage}
        \let\@evenhead\@oddhead \def\@oddfoot{} \def\@evenfoot{} }

\def\numberbysection{\@addtoreset{equation}{section}
        \def\theequation{\thesection.\arabic{equation}}}

\def\underline#1{\relax\ifmmode\@@underline#1\else
        $\@@underline{\hbox{#1}}$\relax\fi}

\def\titlepage{\@restonecolfalse\if@twocolumn\@restonecoltrue\onecolumn
     \else \newpage \fi \thispagestyle{empty}\c@page\z@
        \def\thefootnote{\fnsymbol{footnote}} }

\def\endtitlepage{\if@restonecol\twocolumn \else \newpage \fi
        \def\thefootnote{\arabic{footnote}}
        \setcounter{footnote}{0}} 

\catcode`@=12
\relax

\def\figcap{\section*{Figure Captions\markboth
        {FIGURECAPTIONS}{FIGURECAPTIONS}}\list
        {Figure \arabic{enumi}:\hfill}{\settowidth\labelwidth{Figure
999:}
        \leftmargin\labelwidth
        \advance\leftmargin\labelsep\usecounter{enumi}}}
 \relax
\def\tablecap{\section*{Table Captions\markboth
        {TABLECAPTIONS}{TABLECAPTIONS}}\list
        {Table \arabic{enumi}:\hfill}{\settowidth\labelwidth{Table
999:}
        \leftmargin\labelwidth
        \advance\leftmargin\labelsep\usecounter{enumi}}}
 \relax
\def\reflist{\section*{References\markboth
        {REFLIST}{REFLIST}}\list
        {[\arabic{enumi}]\hfill}{\settowidth\labelwidth{[999]}
        \leftmargin\labelwidth
        \advance\leftmargin\labelsep\usecounter{enumi}}}
 \relax
\makeatletter
\newcounter{pubctr}
\def\publist{\@ifnextchar[{\@publist}{\@@publist}}
\def\@publist[#1]{\list
        {[\arabic{pubctr}]\hfill}{\settowidth\labelwidth{[999]}
        \leftmargin\labelwidth
        \advance\leftmargin\labelsep
        \@nmbrlisttrue\def\@listctr{pubctr}
        \setcounter{pubctr}{#1}\addtocounter{pubctr}{-1}}}
\def\@@publist{\list
        {[\arabic{pubctr}]\hfill}{\settowidth\labelwidth{[999]}
        \leftmargin\labelwidth
        \advance\leftmargin\labelsep
        \@nmbrlisttrue\def\@listctr{pubctr}}}
 \relax
\makeatother

\def\ba{\begin{equation}}
\def\ea{\end{equation}}

\def\a{\alpha}

\def\b{\beta}

\def\g{\gamma}

\def\d{\delta}

\def\e{\epsilon}

\def\no{\noindent}

\def\IR{\relax{\rm I\kern-.18em R}}

\begin{document}


\renewcommand{\theequation}{\thesection.\arabic{equation}}
\csname @addtoreset\endcsname{equation}{section}

\newcommand{\eqn}[1]{(\ref{#1})}
\newcommand{\be}{\begin{eqnarray}}
\newcommand{\ee}{\end{eqnarray}}
\newcommand{\non}{\nonumber}
\begin{titlepage}
\strut\hfill
\begin{center}

\vskip -1 cm


\vskip 2 cm

{\Large \bf Higgsless EW gauge-boson masses from  K-K geometry with torsion}

\vskip .9 in

{\bf Nikolaos A. Batakis}

\vskip 0.2in

Department of Physics, University of Ioannina, \\
45110 Ioannina,  Greece\\
{\footnotesize{\tt nbatakis@uoi.gr}}\\

\end{center}

\vskip .4in

\centerline{\bf Abstract}

\no
In the classical
Higgsless $M^4\times S^3\times S^1$ Kaluza-Klein gauge-theory vacuum,
torsion-induced loss of the  $SU(2)\times U(1)$ left-invariance 
generates a gauge-boson mass term,
in co-existence with the gauge-field kinetic term in the vacuum Hilbert-Einstein action. This
is compulsory in the sense that having one of these terms without the other would mean
violation of basic theorems on holonomy. The effect could be lost only at the very high
massless-EW energy scale, where the ground-state manifold approach for the gauge-theory
vacuum would be inapplicable. Created, as they are, from and as part of pure geometry, these
gauge-boson masses reproduce precisely the spectum of the well-known experimental result.
The apparently exploitable geometric elegance
and the ensuing fundamentally geometric origin of mass may provide new theoretical and
experimental perspectives around and well above the EW scale.

\vfill
\no


\end{titlepage}
\vfill
\eject
\section{Introduction}

In the wake of flow  of LHC results, optimism and doubt mix
as the Higgs sector of the Glashow-Weinberg-Salam model 
for the EW interaction  \cite{gws} undergoes decisive tests
and exploration. The main task is, of course, a formidable one, 
for all the known and spotlighted reasons;
if confirmed, it will mark one more
triumph in the best tradition of progress in physics.
All the same, it is also a precarious one, because the Higgs 
sector, fundamental as well as elusive as it has turned out to be, 
has been founded on an unusually compromising mixture of arbitrariness and inelegance. 
This view may raise objections  but, in any case, it  accords with the wider one on the need 
for updated alternatives. 
If we restrict ourselves, as we will, to the Higgs mechanism, 
alternatives thereoff have apparently been relatively few 
following its collective formulation in the early 60's. This mechanism,  
from the initial status of a \mbox{`clever artifact'}  at the time [presumably to be replaced
by the then expected proper theoretical development] 
has since been elevated to a fundamental part of the
best theory we have for the EW interaction. If the former is shaken, the latter (and not only)
will be in need of even more fundamental reform.

In the present work, the crucial step is to minimally lift the constraint
of  vanishing torsion\footnote
{The notion of torsion is 
here being utilized with its purely geometric
interpretation, namely as introduced by Cartan and exemplified by, e.g., Trautman \cite{traut}, 
{\it not necessarily} in the context of the so-called metric-affine or Einstein-Cartan
theories of gravity.} in a minimally adopted classical Higgs-less $M^4\times S^3\times S^1$
 Kaluza-Klein $SU(2)\times U(1)$ gauge-theory vacuum \cite{duff}.
We will only employ purely geometric quantities 
(no energy-momentum or other such fields of any kind will be introduced)
and we will proceed  by fundamental considerations in Physics, as guided by general
principles  in  Riemannian and group-theoretic arguments from holonomy in Riemann-Cartan
 differential geometry \cite{helga},\cite{traut}.

\no

\section{Basic setting and an overview}

With some sacrifice of brevity for clarity, 
we will now introduce notation for 
the already mentioned   8-dimensional Kaluza-Klein vacuum
\be
M^8 &=& M^4\times S^3\times S^1\ ,
\label{m8}
\ee
with holonomic coordinates, say, $x^\mu$ in $M^4$ and $y^{\hat{m}}$  in  $S^3\times S^1$.
As we will mostly concentrate on the latter, it is advantageously  equivalent to employ, as
 we will, non-holonomic coordinates,
implicitly defined by an invariant basis of 1-forms $\ell^{\hat{a}}$. 
We first note that,
strictly speaking, the definition  (\ref{m8}) holds globally only for the ground-state 
$M^8$ direct-product manifold.
For the menagerie of  $M^8$  manifolds to which one can extend beyond  the ground-state $M^8$, 
the definition  (\ref{m8}) holds only locally, namely only in open neighbourhoods, 
as defined by, e.g., Cartan frames $e^A$ (namely {\em  local} non-holonomic orthonormal bases 
of 1-forms), such as the ones we will employ. 
In any case, as the definition (\ref{m8}) is understood to define manifolds (rather than just bare topology), 
it must be supplemented by whatever pair of metric $g_{MN}$ plus  
(e.g., Christoffel) connection  $\gamma^M_{\;\;NP}$ 
or, equivalently, by whatever pair of frames $e^A$  plus 1-form connection $\omega^A_{\;\;B}$ is being used. 
This, if needed for clarity, will be shown explicitly as $M^8(e^A,{\omega}^A_{\;\;B})$, or 
$M^8(\omega)$ for short, and understood along with its whatever 8-beins $e^A_M$ and inverse $E^M_A$.
All such $M^8$ manifolds will  
automatically have the same components for their metric  $\eta_{AB}$, by definition identical 
to their common signature, namely 
 \be
\eta_{AB} = {\rm diag} (-1,+1,+1,+1;+1,+1,+1,+1)\ .
\label{metric}
\ee 
The range for the world indices $A,B,\ldots$ is 
$A\!=\!(\alpha;\hat{a})$, with $\alpha\!=\!(0,1,2,3)$, 
$\hat{a}\!=\!(a,4)$, $a\!=\!(1,2,3)$, and the same for 
coordinate (usually holonomic)  indices $M,N,\ldots$ with
$M\!=\!(\mu;\hat{m})$, etc. $SU(2)\times U(1)$ group indices
I,J,\ldots have the same range as indices in the $S^3\times S^1$ manifold,
namely $I\!=\!(i,4)$, $\!i=\!(1,2,3)$.
The duality between  $e^A$ and the $E_B$ basis  for tangent-vectors in $M^8$ relates 
(\ref{metric}) to the conventional {\it holonomic-coordinate} metric  $g_{M\!N}$ as
\be
 e^A(E_B)= e^A_ME^M_B = \delta^A_B :\;\;\;\;\;\;\;\;\;\;\eta_{AB}=
E^M_A E^N_B  g_{MN} \;\; \Longleftrightarrow\;\; g_{MN}=e^A_Me^B_N \eta_{AB}\ . 
\label{holmetric}
\ee
In the case of the ground-state manifold $M^8(\stackrel{o}{e^A},{\gamma^A}_B)$,  
or $M^8(0)$ for short, the latter reduces to its familiar block-diagonal expression as
\be
\stackrel{o}{e}{}^A_M=\left(  \begin{array}{c} 
{\stackrel{o}{e}}{}^{\alpha}_{\mu}\\ 
{\stackrel{o}{e}}{}^{\hat{a}}_{\hat{m}}
\end{array} \right)      
 \;\;\; \Longleftrightarrow\;\;\;
 {\stackrel{o}{g}}_{MN}=\left(  \begin{array}{cc} \stackrel{o}{g}_{\mu\nu} & 0 \\ 
0 & \stackrel{o}{g}_{\hat{m}\hat{n}}
\end{array} \right)    \ .
\label{grmm}
\ee
The set of Christoffel connection 1-forms $\gamma^A_{\;\;B}$
needs no special mention as it  
is, of course, always uniquely determined by the metric or the frames.
When we lift the zero-torsion constraint 
(but keep the wider zero-metricity constraint),
the general connection is
\be
\omega^A_{\;\;B} &=& \gamma^A_{\;\;B} +K^A_{\;\;B} \ ,
\label{conect}
\ee
where the ${\gamma}^A_{\;\;B}$ set of 1-forms (which is not a tensor as it remains a 
Christoffel connection) is antisymmetric in $A,B$  if expressed  in the Cartan frames we employ. 
The contorsion $K^A{\!}_B$ is the standard antisymmetric in $A,B$ tensor-valued 1-form, 
with components related algebraically to those of torsion. 
Cartan's structure equations
\be
{\cal T}^A:&=&{\cal D} e^A:=de^A+\omega^A{\!}_B\wedge e^B =De^A+K^A{\!}_Be^B = K^A{\!}_Be^B=\frac{1}{2}{\cal T}^A_{\;\;EP}e^E\wedge e^P,  \
\label{torsion}
\\
{\cal R}^A_{\;\;B}:&=& d\omega^A_{\;\;B}+\omega^A_{\;\;E}\wedge\omega^E_{\;\;B}=\frac{1}{2}{\cal R}^A_{\;\;BEP}e^E\wedge e^P,
\ 
\label{riemann}
\ee
which also involve the covariant exterior derivatives $ {\cal D}$ and $D$ w.r.t. 
$\omega^A_{\;\;B}$ and $\gamma^A_{\;\;B}$ with $De^A\equiv 0$, define the fundamental pair of
the torsion ${\cal T}^A$ and the Riemann-curvature  ${\cal R}^A_{\;\;B}$ 2-forms.
The trace of the former is identical (up to sign) to that of the contorsion 
$ K^A{\!}_B$; the traces of the Riemann, through the Ricci tensor, produce
the curvature scalar $R$. This allows us to write down the full content of the Hilbert-Einstein action as 
\be
{\cal I}_{H-E} {\;\;}\sim & \int _{M^8}{\cal R}^A_{\;\;B}\wedge{\varepsilon}_A^{\;\;B} = \int _{M^8}R {\;} \varepsilon \ ,
\label{hilbein}
\ee
with
the  standard definitions for the ${M^8}$ 6-form density and volume elements
\be
\varepsilon_{A_1A_2}:=\frac{1}{(8-2)!}\;\; \epsilon_{A_1A_2\dots A_8}\;e^{A_3}\wedge \dots \wedge e^{A_8},\;\;
\varepsilon: =\frac{1}{8!}\;\; \epsilon_{A_1A_2\dots A_8}\;e^{A_1}\wedge \dots \wedge e^{A_8} \ .
\label{eta}
\ee

We will now proceed with a  preliminary and informal  overview of our results. 
All geometric and symmetry aspects (to be properly examined in section 4) are involved in  
three basic manifolds, related as they evolve from first to last as
\be
M^8(0)\!:=\!M^8(\stackrel{o}{e^A},{\gamma}^A_{\;\;B})\;  
\stackrel{[\xi\!\cdot\!{\cal{A}}]}{\longrightarrow}\;\;\;M^8(\gamma)\!:=\!M^8(e^A,{\gamma}^A_{\;\;B}) 
\;\stackrel{{\rm torsion}}{\longrightarrow}\;\;\;
M^8(\omega)\!:=\!M^8(e^A,{\omega}^A_{\;\;B}).\ 
\label{evolv}
\ee  
In standard non-abelian Kaluza-Klein theory, the first step in (\ref{evolv}) is 
realized with the
$\!SU\!(2)\!\!\times\!\! U\!(1)\!$ gauge potentials $\cal{A}$, along with the  
$S^3\times S^1$ Killing vectors $\xi_I$,  introduced in products as off-diagonal 
elements in the metric (\ref{grmm}).
We'll do the same thing here with the employment of the scaleless (1,1)-rank `diagonal'
 tensor $\big{[}\xi\!\cdot\!{\cal{A}}\big{]}$ with components
\be
\big{[}\xi\!\!\cdot\!\!{\cal{A}}\big{]}^{\hat{a}}_\beta:=
\xi_i^{\hat{a}}{\cal{A}}^i_\beta\sin{\theta}+\xi_4^{\hat{a}}{\cal{A}}^4_\beta\cos{\theta} ,
  \  
\label{dotc}
\ee
to be formally introduced later on. 
As it happens, in  the first step in (\ref{evolv}),  $\big{[}\xi\!\cdot\!{\cal{A}}\big{]}$
{\em tilts} (occasionally better visualized as `shakes-up')  the frames with 
point-depended translations and rotations in ${M^8}$, 
but under the zero-torsion constraint for the connection; 
with the second step in (\ref{evolv}), the frames remain the same but the constraint  is lifted and torsion emerges,
of course in the context of Riemann-Cartan  geometry. 

Turning from geometric to symmetry aspects,
of particular interest are the $S^3$ sections (slicings) of the $S^3\times S^1$ torus in (\ref{m8}).
In the initial ground-state manifold, 
this $S^3$  is a {\em round} one.\footnote{Round, aka maximally symmetric, aka homogeneous and isotropic, 
aka invariant under the
translations and rotations of its full isometry group of motions, aka invariant under
 the $\;SU\!(2)\!\times\!SU\!(2)\;$ left and right action of $SU(2)$ on  its group manifold $S^3$.} 
With  the first step in (\ref{evolv}),
the round $S^3$  loses `half' of its symmetry, now reducing to 
a {\it squashed} $S^3$, which is only homogeneous, with only its left-invariant 1-forms $\ell ^a$
 surviving. 
With the second step in (\ref{evolv}), as the zero-torsion constraint is lifted,
this squashed $S^3$ 
looses all of the remaining symmetry, practically stripped-down to almost bare topology, 
in its terminal reduction to a {\it deformed} $S^3$.  However, any sense of degeneration
would be false at this point, because what has actually taken place 
is a rather miraculous phenomenon of {\em generation}, namely that of mass from the vacuum; 
the loss of the last symmetry is directly responsible for the simultaneous emergence
of the  $\!SU\!(2)\!\times\! U\!(1)\!$  gauge-boson  masses, thus created  from and as part of pure geometry.

\no

\section{Relating to standard K-K $SU\!(2)\!\!\times\!\!U\!(1)$ gauge theory}

We begin by first relating our  preceding overview to more rigorous notation. The left-invariant 
non-holonomic set of 1-forms $\ell^a$ on $S^3$ can be supplemented with an extra 1-form  $\ell^4$ for $U(1)$,
 hence with  $d\ell^4=0$, 
thus enlarged to a left-invariant basis  $\ell^{\hat{a}}$ on
$S^3\times S^1$, with dual $L_{\hat{a}}$. Their non-vanishing Maurer-Cartan equations and commutation 
relations are  
\be
\ell^{\hat{a}}(L_{\hat{b}})=\d^{\hat{a}}_{\hat{b}}\;\;\;\Longrightarrow\;\;\;\;\; d\ell^a=-\frac{1}{2} 
\epsilon^a{\!}_{bc}\ell^b\! \wedge\!\ell^c \;\;\; \Longleftrightarrow\;\;\;
[L_b,L_c]= \epsilon^a{\!}_{bc}{ }L_a \ .
\label{invariant}
\ee
Associated with the group of motions (isometries) on $S^3\times S^1$,
there are four Killing vectors $\xi_I$ which {\em survive} when the round $S^3$  in $M^8(0)$
reduces  to a squashed one in $M^8(\gamma)$, with components $\xi_I^{\hat{a}}$ which
are identical in any one of the $M^8(0)$, $M^8(\gamma)$ or $M^8(\omega)$ manifolds (a property we will
 shortly expand on in this section).
Their non-vanishing  commutation relations are
\be
{\cal{L}}_{\xi_j}\xi_k:=[\xi_j,\xi_k]= \epsilon^i{\!}_{jk}{ }\xi_i \ ,
\label{killingcom}
\ee
where ${\cal{L}}_{\xi_I}$ is the Lie derivative for each  Killing vector $\xi_I$. 
Their orthogonality and  lengths are fixed by the {\em slicing angle} $\theta$, with values in $(0,\pi/2)$, 
as\footnote{
The slicing angle $\theta$ should be carefully distinguished from Weinberg's {\it mixing angle} $\theta_W$. 
Although fundamentally distinct, $\theta$ and $\theta_W$ 
will relate to one-another in a subtle way, apparently nebulous in the literature, as it will emerge in the 
sequel.}
\be
\xi_I^{\hat{a}}\xi_J^{\hat{b}}\delta_{{\hat{a}}{\hat{b}}}:= 
\Big{(} \frac{{\rm L}_o}{\sin{\theta}} \Big{)}^2\delta_{ij}\delta^i_I\delta^j_J+ \Big{(} 
\frac{{\rm L}_o}{\cos{\theta}} \Big{)}^2\delta_{44}\delta^4_I\delta^4_J \ .
\label{killinglengths}
\ee
The scale ${\rm L}_o$ of the components is  imposed by the frames and the lengths  ${\rm L}_o/\sin{\theta}$ 
 and ${\rm L}_o/\cos{\theta}$ are 
proportional to the radii of $S^3$ and $S^1$ in the particular slicing of the $S^3\times S^1$ torus,
 as fixed by $\theta$.
The $\xi_I$ provide a  basis
for tangent vectors in $S^3\times S^1$, just like the earlier introduced $L_{\hat{a}}$.
However, while the  $L_{\hat{a}}$  are {\em ab initio} left-invariant,
a fact equivalently expressed as\footnote{
Since the frames $e^{\hat{a}}$ will be constructed from the invariant basis $\ell^{\hat{a}}$, either of
 the two equations in (\ref{killonell}) expresses precisely the content of the related Killing equations.}
\be
{\cal{L}}_{\xi_I}\ell^{\hat{a}}=0,\;\;\;{\cal{L}}_{\xi_I}L_{\hat{a}}:=[\xi_I,L_{\hat{a}}]=0 \ ,
\label{killonell}
\ee
the $\xi_I$ cannot possibly form a left-invariant basis, in view of (\ref{killingcom}). 
Therefore, under ordinary circumstances, the $\xi_I$ would be an odd and cumbersome (albeit fully 
legitimate) basis to employ in left-invariant environments, 
such as those involving the round or even the squashed   $S^3{\;}$'s. This observation will be useful
 to us later on.

We now proceed with the formal definition of the first step in  (\ref{evolv}), which starts 
with the frames $\stackrel{o}{e^A}$  and dual $\stackrel{o}{E_{\hat{b}}}$ in $M^8(0)$, introduced as columns and lines, with
\be
\stackrel{o}{e^A}\!\!&:&\;\; \big{[}\!\stackrel{o}{e^\a}=\stackrel{o}{e^\a}_\mu\!dx^\mu\;;
\;\;\stackrel{o}{e^{\hat{a}}}=\!\!{\rm L}_o({\rm l}_3\d^{\hat{a}}_a\ell^a\!+
\!{\rm l}_1\d^{\hat{a}}_4\ell^4)\big{]},\nonumber \\
\stackrel{o}{E_B}\!\!&:&\;\; \big{(}\!\!\stackrel{o}{E_\b}=\!\stackrel{o}{E_\b^\mu}\!\!\partial_\mu\;
;\;\;\stackrel{o}{E_{\hat{b}}}=\!{\rm L}_o^{-1}({\rm l}_3^{-1}\d_{\hat{b}}^bL_b\!+
\!{\rm l}_1^{-1}\d_{\hat{b}}^4L_4)\big{)}     
  \ ,
\label{eo}
\ee
where the scaleless parameters ${\rm l}_3, {\rm l}_1$ fix the radii of
 $S^3$ and $S^1$  in ${\rm L}_o$-length units.
With the first step in (\ref{evolv}), the above frames evolve to
\be
\stackrel{o}{e^A}\;\rightarrow\;\; {e^A}:={ }\stackrel{o}{e^A}+
g\big{[}\xi\!\!\cdot\!\!{\cal{A}}\big{]}^{\hat{a}}\delta_{\hat{a}}^A
 \;\;\; \Longleftrightarrow\;\;\;
\stackrel{o}{E_B}\;\rightarrow\;\; {E_B}={ }\stackrel{o}{E_B}-
g\big{[}\xi\!\!\cdot\!\!{\cal{A}}\big{]}_{\beta}\delta^{\beta}_B
  \ ,
\label{tilt}
\ee
with  $E_B$ following by duality and with $g$ a scaleless coupling parameter.\footnote
{We recall that the basic scales, like ${\rm L}_o$ (to be thought-of as Planck scale),
are carried by the frames. As 
a rule, all other geometric quantities  employed must have derivable scale or be
 scaleless as, e.g., with all entries in (\ref{conect})
and  (\ref{killonell}). In the scaleless coupling parameter $g=\sqrt{2}\kappa/{\rm L}_o$,
 the denominator will de-scale 
$\xi_I^\alpha$, circumstancially scaled by ${\rm L}_o$  in (\ref{dot}), and  then
 $\kappa^2=8\pi G_N$
(to be thought-of as the gravitational coupling) will
provide the missing scale. Variant reasoning will later reveal the EW-scale parameter ${\rm L}_1$.} 
We have already mentioned  $ \big{[}\xi\!\!\cdot\!\!{\cal{A}}\big{]}$, 
which is a mixed (1,1)-rank tensor with components 
\be
\big{[}\xi\!\!\cdot\!\!{\cal{A}}\big{]}^A_B:=  
\big{[}\xi\!\!\cdot\!\!{\cal{A}}\big{]}^{\hat{a}}_{\beta}\delta_{\hat{a}}^A \delta^{\beta}_B
  \ .
\label{ttcomp}
\ee
These can be explicitly specified by
the (1,0)-valued 1-form and (0,1)-valued tangent vector in (\ref{tilt}), defined and calculated as
\be
\big{[}\xi\!\!\cdot\!\!{\cal{A}}\big{]}^{\hat{a}}:\!\!\!&=&
\xi_i^{\hat{a}}{\cal{A}}^i\sin{\theta}+\xi_4^{\hat{a}}{
\cal{A}}^4\cos{\theta} ,
\nonumber \\
\big{[}\xi\!\!\cdot\!\!{\cal{A}}\big{]}_{\beta}&=&\xi_i{\cal{A}}^i_{\beta}\sin{\theta}+\xi_4{\cal{A}}^4_{\beta}\cos{\theta} 
  \ . 
\label{dot}
\ee
We observe that, by moving  $g\big{[}\xi\!\cdot\!{\cal{A}}\big{]}$ on the other side of each
 equation  in (\ref{tilt}),
the latter can be made to define  $\stackrel{o}{e^A}$   
in terms of ${e^A}$ and $ \big{[}\xi\!\cdot\!{\cal{A}}\big{]}$; the question then arises as to
 which frame are the (\ref{ttcomp})
components expressed in. Actually, and this is part of the elegance of the K-K scheme, there is
 no obscurity or inconsistency involved.
The reason is that 
$\big{[}\xi\!\cdot\!{\cal{A}}\big{]}$ belongs to a class of geometric quantities that have the 
{\it same} components 
in any one of the $M^8(0)$, $M^8(\gamma)$ or $M^8(\omega)$ manifold frames, just like the 
already-mentioned $\xi_I^{\hat{b}}$. The
result is due to two frame and vierbein  invariances in  (\ref{tilt}), as stated below, which
 give rise to that special class, with several examples 
pointed-to by the long arrow
\be
\stackrel{o}{e^\alpha}\rightarrow e^\alpha\!\!=\stackrel{o}{e^\alpha},\; \stackrel{o}{E_{\hat{a}}}
\rightarrow E_{\hat{a}}\!=\stackrel{o}{E_{\hat{a}}}\;
\Longrightarrow \;
e^{\alpha}_\mu, \; E_{\hat{b}}^{\hat{m}}, \; L_{\hat{b}}^{\hat{m}}, \;\xi_I^{\hat{b}}, \;{\cal{A}}_\b^I,
\;\big{[}\xi\!\cdot\!{\cal{A}}\big{]}^{\hat{a}}_{\b}, \;\varepsilon_{A_1\cdots A_8}, \partial_y. 
  \ 
\label{ttcompo}
\ee
Examples of typical behavior or counter-examples to the above involve differing components, e.g., 
unlike $\partial_y$ in (\ref{ttcompo}) and contrasted to it, we have 
\be
\stackrel{o}{E_\b}
\rightarrow E_\b\!=\stackrel{o}{E_\b}-\big{[}\xi\!\!\cdot\!\!{\cal{A}}\big{]}_{\beta}
\Longrightarrow \;\;\;\partial_\mu
\rightarrow\;\;\partial_\mu-g(\xi_i{\cal{A}}^i_{\mu}\sin{\theta}+\xi_4{\cal{A}}^4_{\mu}\cos{\theta})
  \ ,
\label{ttcom}
\ee
expressed  in holonomic coordinates too,  with $\partial_\mu\!:=\partial/ \partial x^\mu$. The latter 
supplies us with a generalized minimal-coupling prescription, here obtained as a
 rigorous result from (\ref{tilt}).
On the same grounds, and in view of the $\stackrel{o}{\varepsilon}=\!\varepsilon$ 
result in (\ref{ttcompo})  for all ${M^8}$ volume elements, the Hilbert-Einstein 
action (\ref{hilbein}) will also evolve 
along the (\ref{evolv}) sequence, as depicted  in
\be
{\cal I}_{HE}(0)\sim\!\!\int _{M^8(0)}\!\!R(0)\varepsilon\;
\stackrel{[\xi\!\cdot\!{\cal{A}}]}{\longrightarrow}\;\;\;
{\cal I}_{HE}(\gamma)\sim\!\!\int _{M^8(\g )}\!\!R(\gamma)
\varepsilon\;\stackrel{{\rm torsion}}{\longrightarrow}\;\;\;
 {\cal I}_{HE}(\omega)\sim\!\!\int _{M^8(\omega)}\!\!R(\omega)\varepsilon,
\ 
\label{evolvact}
\ee
where $R(0)$, $R(\gamma)$ and $R(\omega)$ are the curvature scalars in the respective manifolds.

\no

\section{Torsion as prescribed by geometry and symmetry}

Here we will essentially review more or less known results in standard 
Kaluza-klein $\;SU\!(2)\!\times\!U\!(1)\!$ gauge theory, 
and recall some results from holonomy (as they relate to our present needs and discussed in the last section); 
the aim is to explicitly deduce our minimally adequate torsion from
 fundamental geometric and symmetry aspects, 
before we proceed with our main results in the next section. 
The  Lagrangians ${\cal L}_{HE}$, which can be read-off from the actions 
(\ref{evolvact}) as proportional to the corresponding curvature scalars  $R$, are directly
calculable from the 
geometric definitions (\ref{conect}-\ref{hilbein}).
With the usual adjustments,\footnote{With hardly any non-trivial exception,
the curvature scalars $R$ contain surface terms
{\em omitted} in the corresponding Lagrangians.
However, the latter differ from $R$ also with {\em added} terms, notably 
effective cosmological constants coming from reduction to 4 spacetime dimensions. 
Both adjustment are here understood as present in the ground state 
Lagrangian ${\cal L}_{HE}(0)$.} 
we obtain the well-known result for  $M^8(\gamma)$ as
\be
{\cal L}_{HE}(\gamma)={\cal L}_{HE}(0)-\frac{1}{2}\kappa^2{\cal F}^2
 \ .
\label{lagr}
\ee
The calculation may begin with the defining setup in the previous section for 
the basic preliminary result
\be
de^{\hat{a}}=g(\sin{\theta}\xi^{\hat{a}}_i{\cal F}^i+\cos{\theta}\xi^{\hat{a}}_4
{\cal F}^4)-\frac{1}{2{\rm l}_3{\rm L}_o}\d^{\hat{a}}_a\e^a_{\;bp}e^b\wedge e^p 
 \ .
\label{eF}
\ee
The field strengths therein have been introduced as
\be
{\cal F}^I:=d{\cal A}^I+\frac{1}{2}g\sin{\theta} \;\d_i^I\e^i_{jk} {\cal A}^j
\wedge{\cal A}^k
 \ ,
\label{F}
\ee
in an obviously left-invariant environment, leading to (\ref{lagr}) with
\be
{\cal L}_{\xi_I}{\cal F}^J=0,\;\;\;{\cal L}_{\xi_I} {\cal L}_{HE}(\g)=0
 \ .
\label{LL}
\ee
To repeat the calculation for ${\cal L}_{HE}(\omega)$, we observe that the
 connection $\omega$ in (\ref{conect}) for $M^8(\omega)$
 involves the additional contribution from the contorsion tensor $ K^A_{\;\;B}$. 
Without any restriction for the latter,
the result of this calculation  is
\be
 {\cal L}_{HE}(\omega)={\cal L}_{HE}(0)-\frac{1}{2}\kappa^2{\cal F}^2 + K^{APB}K_{BPA}-({\rm tr}K)^2
 \ .
\label{lagt}
\ee
We still have a left-invariant ${\cal F}^I$, with the same definition ($\ref{F}$), 
and with the first equation in ($\ref{LL}$) still valid. 
However, as we expect, this cannot be the case for the contorsion tensor $ K^A_{\;\;B}$, 
hence it will no-longer be the case for the Lagrangian ${\cal L}_{HE}(\omega)$ either.

The result from holonomy (to be briefly reviewed in the last section, as mentioned) is that  
the torsion tensor ${\cal T}^A$,  as it emerges in the present context,  will have to
 be proportional to the gauge potentials  ${\cal A}^I$;
we thus need to clarify whether the latter should enter the specification of  ${\cal T}^A$
as $\big{[}\xi\!\!\cdot\!\!{\cal{A}}\big{]}$
or as  $\xi_I{\cal{A}}^I$. The choice is already rooted in the contemplation following  (\ref{killonell}), 
as to why a non-invariant (albeit legitimate) basis $\xi_I$ is employed in the left-invariant environment of $M^8(\g)$. 
In fact, to just  arrive  at  (\ref{lagr}) or even  (\ref{lagt}), we could have entirely dispensed with 
the slicing-angle $\theta$  
{\em and} the Killing vectors, altogether. This
is already quite obvious from (\ref{LL}) and the trivially redundant presence of $\theta$ in (\ref{F}).
It can also be explicitly demonstrated  with the employment of the left-invariant basis,
easily effected with the symmetric alternative
$L_o\delta_I^{\hat{a}}$ replacing  $\xi_I^{\hat{a}}$ in  $\big{[}\xi\!\!\cdot\!\!{\cal{A}}\big{]}$,
formally with
\be
\big{[}\xi\!\!\cdot\!\!{\cal{A}}\big{]}^{\hat{a}}\rightarrow\;\;\;\big{[}{\rm I}\!\cdot\!{\cal A}
\big{]}^{\hat{a}}:= L_o\delta_I^{\hat{a}}{\cal {A}}^I
  \  
\label{dot1}
\ee
in (\ref{tilt}).
The calculation, now significantly simplified
in the explicitly symmetric environment, leads (modulo trivial re-definitions) 
to precisely the same Lagrangians  (\ref{lagr}) and  (\ref{lagt}).
The compatibility of the use of two fundamentally different frames, namely the non-invariant $\xi_I$ and 
the left-invariant $L_{\hat{a}}$ (or, equivalently, the  $E_{\hat{a}}$ in (\ref{lagr}))
is achieved as follows: The particular choice of dependence of $\big{[}\xi\!\cdot\!{\cal{A}}\big{]}$ 
on ${\theta}$  in (\ref{dot})
is made in conjunction with the {\mbox{\em ad hoc}} choice of the radii
${\rm L}_o/\!\sin{\theta}$  and ${\rm L}_o/\!\cos{\theta}$  in the (\ref{killinglengths}) slicing,
with the intention to precisely cancel out
the $\theta$-dependence.  
This renders any $\theta$-slicing of the $S^3\times S^1$  torus
as good as any other, therefore it {\em explicitly} re-establishes  left-invariance, which was actually never lost.
However, all
this indeed redundant involvement of $\theta$-slicing is not only {\it not} useless, 
it is, in fact, crucial and  irreplaceable, as we will see in a moment. 
We now let a mixing-angle $\theta_W$  introduce through $ K^A_{\;\;B}$ a randomly occurring 
symmetry-breaking direction, say $\Xi^{\hat{a}}$,
tangent to the $S^3\times S^1$ torus in $M^8(\omega)$. With  proper normalization to unit length
 and without loss of generality, we follow the
conventional choice in context, with
\be
\Xi^{\hat{a}}:=\frac{1}{\sqrt{2}{\rm L}_o}(\xi_3^{\hat{a}}\sin{\theta_W} +\xi_4^{\hat{a}}\cos{\theta_W}) \ ,
\label{xi}
\ee
inducing, correspondingly, the conventional mixing of $SU\!(2)\!\!\times\!\!U\!(1)$ gauge potentials.
The $S^3$ sections, now identified by the $\theta=\theta_W$ 
slicings,\footnote{Here,
the subtle interrelation between $\theta$ and $\theta_W$ we referred-to earlier is rather clear.
By being {\em \mbox{a posteriori}} set equal to $\theta_W$, $\theta$ itself overrides its until-then
 redundant presence, while $\theta_W$ acquires an additional, now `slicing'
property, which it couldn't have otherwise. However, with $\theta=\theta_W$, the until-then valid 
`any $\theta$ slicing is as good as any other' is obviously lost,
subject to the same symmetry breaking as induced by (\ref{xi}).
Of course, in the absence of symmetry breaking, $\theta$ is trivially redundant and $\theta_W$ is trivially irrelevant.}
 are precisely the
deformed $S^3$'s at every point in $M^8(\omega)$, as already described in section 2.

We now have two results at hand, first to opt for 
$\xi_I{\cal{A}}^I$ rather than  $\big{[}\xi\!\!\cdot\!\!{\cal{A}}\big{]}$
(in view of the circumstancially  {\mbox{\em ad hoc}} symmetry-restoring r\^{o}le of the latter),
and second the symmetry-breaking $\Xi^{\hat{a}}$  vector in (\ref{xi}), 
as they complement the mentioned
holonomy-group requirement of 
proportionality of the torsion tensor ${\cal T}^A$ to ${\cal{A}}^I$.\footnote
{Viewed as a vector-valued form in $M^8(\omega)$, ${\cal T}^A$  must be proportional to the gauge potentials 
not just numerically but essentially to their vectorial directions $\d^{\a\b}{\cal{A}}^I_\b$, 
tangent to the $M^4$ subspaces of $M^8(\omega)$.} 
The result (essentially unique by minimality)  for our torsion dictates
{\em only} ${\cal T}^\a$ components,  properly scaled as\footnote
{The components of ${\cal T}^\a$ carry a scale from whatever frames employed, here ${\rm L}_o$ from $e^{\hat{a}}$, 
exactly as the dimensionless ${\cal F}$ does from $e^{\a}$, but,
unlike the latter, the {\em dimensionful} $\;{\cal T}^\a$  requires an extra frame-independent length-scale, 
here supplied by 
${\rm L}_1$. A more rigorous version of this claim follows from Cartan's first structure equation in (\ref{torsion}),
which quantifies the enlargement of geometry with torsion.} 
\be
{\cal T}^{\a}\sim \frac{g}{{\rm L}_1}\d^{\a\b}\Xi^{[\hat{a}}\xi_I^{\hat{b}]}{\cal{A}}^I_\b \ .
\label{t}
\ee
In the next section we turn to our main results, presented  in detail
sufficient for their reproduction. 

\no

\section{$SU\!(2)\!\!\times\!\!U\!(1)$ gauge-boson masses from torsion}

Conforming to standard notation in anticipation of the presence of a gauge-boson mass term (with the correct sign!)
 in the Larangian  (\ref{lagt}),
we may re-write it as
\be
 {\cal L}_{HE}(\omega)={\cal L}_{HE}(0) - \frac{\kappa^2}{2}{\cal F}^2-\kappa^2M_{IJ}{\cal{A}}^I_\a{\cal{A}}^J_\b\d^{\a\b}
 \ ,
\label{lagm}
\ee
wherefrom, observing that the torsion in (\ref{t}) is traceless (therefore, so is the contorsion it produces)
 we reed off (\ref{lagt}) 
\be
\kappa^2M_{IJ}{\cal{A}}^I_\a{\cal{A}}^J_\b\d^{\a\b}= - K^{APB}K_{BPA}
 \ .
\label{mk}
\ee
In the context of our definitions, the components  of the torsion and contorsion tensors are interrelated as
\be
 K_{ABP}=-\frac{1}{2}\big{(}T_{ABP}+T_{BPA}-T_{PAB}\big{)}\;\;\;\;\;\Longleftrightarrow\;\;\;\;\;T_{ABP}=-K_{ABP}+K_{APB}
 \ ,
\label{kt}
\ee
so our basic result in (\ref{t})  supplies  for our torsion and contorsion tensors  their  only
 non-vanishing components as
\be
 -K^{\a\hat{b}\hat{p}}=+K^{\hat{b}\a\hat{p}}=+K^{\hat{b}\hat{p}\a}=+\frac{1}{2}T^{\a\hat{b}\hat{p}}=\frac{g}{{\rm L}_1}\d^{\a\b}\Xi^{[\hat{b}}\xi_I^{\hat{p}]}{\cal{A}}^I_\b \ .
\label{ktc}
\ee
The straightforward  substitution of (\ref{ktc})  in (\ref{mk})  quantifies the mass matrix as
\be
M_{IJ}=({\rm L}_o{\rm L}_1)^{-2} \big{[}(\Xi)^2\xi_I^{\hat{p}}\xi_J^{\hat{q}}\delta_{{\hat{p}}{\hat{q}}} -
(\Xi^{\hat{p}}\xi_I^{\hat{q}}\delta_{{\hat{p}}{\hat{q}}})
(\Xi^{\hat{r}}\xi_J^{\hat{s}}\delta_{{\hat{r}}{\hat{s}}}) \big{]}
 \ ,
\label{mm}
\ee
where   $(\Xi)^2=1$ and
\be
\Xi^{\hat{p}}\xi_I^{\hat{q}}\delta_{{\hat{p}}{\hat{q}}} =\frac{{\rm L}_o}{\sqrt{2}}
\Big{(}\frac{\sin{\theta_W}}{\sin^2{\theta}}\delta_{3J} + \frac{\cos{\theta_W}}{\cos^2{\theta}}\delta_{4J}\Big{)}
 \ .
\label{xx}
\ee
{\em {After}} setting $\theta=\theta_W$ (as we will do from now on), we introduce for brevity
\be
(\Xi\cdot\xi)_I:=\Xi^{\hat{p}}\xi_I^{\hat{q}}\delta_{{\hat{p}}{\hat{q}}}([\theta\!=\!\theta_W]) =
\frac{{\rm L}_o}{\sqrt{2}}\Big{(}\frac{1}{\sin{\theta_W}}{\delta_{3I}} + \frac{1}{\cos{\theta_W}}\delta_{4J}\Big{)}
 \ ,
\label{xm}
\ee
so, using (\ref{killinglengths}), (\ref{xi}) and (\ref{xm}), we may express (\ref{mm}) as  
\be
M_{IJ}&=&({\rm L}_o{\rm L}_1)^{-2} \big{[}\xi_I^{\hat{p}}\xi_J^{\hat{q}}\delta_{{\hat{p}}{\hat{q}}} -
(\Xi\cdot\xi)_I(\Xi\cdot\xi)_J \big{]} \label{mm1}
 \\&=&({\rm L}_1\sin{\theta_W})^{-2}
 \Big{[}\delta_{ij}\delta^i_I\delta^j_J-\frac{1}{2}\big{(}\delta^3_I\delta^3_J+\tan^2{\theta_W}\delta^4_I\delta^4_J 
+\tan{\theta_W}(\delta^3_I\delta^4_J +\delta^4_I\delta^3_J)\big{)}\Big{]}
 \ ,
\nonumber
\ee
or, in the more conventional  matrix notation,
\be
M_{IJ}=({\rm L}_1\sin{\theta_W})^{-2}
\left(\begin{array}{ccccc} 1&{\;} & 0&0&0 \\ 0&{\;}&1&0&0\\0&{\;}&0&\frac{1}{2}&-\frac{1}{2}
\tan{\theta_W}\\0&{\;}&0&-\frac{1}{2}\tan{\theta_W}&\frac{1}{2}\tan^2{\theta_W}
\end{array} \right)          
 \ .
\label{mm2}
\ee
Either of 
\be
\Delta^I_{\;\;J}=
\left(  \begin{array}{ccccc} 1&{\;} & 0&0&0 \\ 0&{\;}&1&0&0\\0&{\;}&0&-\cos{\theta_W}& 
\sin{\theta_W}\\0&{\;}&0&+\sin{\theta_W}&\cos{\theta_W}
\end{array} \right)  \;{\rm {or}}\;\;
\left(  \begin{array}{cccc} 1/\sqrt{2} & 1/\sqrt{2}&0&0 \\ 
-i/\sqrt{2}&i/\sqrt{2}&0&0\\0&0&-\cos{\theta_W}& \sin{\theta_W}\\0&0&+\sin{\theta_W}&\cos{\theta_W} \end{array} \right)          
 \ 
\label{diag}
\ee
diagonalizes $M_{IJ}$  to its eigenvalues as
\be
M^I_{\;\;J}=({\rm L}_1\sin{\theta_W})^{-2}
\left(  \begin{array}{ccccc} 1&{\;} & 0&0&0 \\ 0&{\;}&1&0&0\\0&{\;}&0&\frac{1}{2}(\cos\theta_W)^{-2}& 0\\0&{\;}&0&0&0
\end{array} \right)=        
\left(  \begin{array}{cccc} m^2_W & 0&0&0 \\ 0&m^2_W&0&0\\0&0&\frac{1}{2}m^2_Z& 0\\0&0&0&0
\end{array} \right)   
 \ . 
\label{diagm}
\ee
The  acquired gauge-boson  masses (with their spectrum
also interpretable as angular `spring-constants' of the anisotropic vacuum, as we will see), are fixed as
\be
 m_W=({\rm L}_1\sin{\theta_W})^{-1}\;,\;\;\;\; m_Z=({\rm L}_1\sin{\theta_W}\cos{\theta_W})^{-1}    
 \ , 
\label{diagmm}
\ee
with the $\rho$ parameter, which is defined as  $\rho:=m^2_W/( m_Z\cos{\theta_W})^2$,  post-dicted 
 directly from (\ref{diagmm}) as  $\rho=1$.
In terms of the usual expressions for the physical gauge bosons, actually read off the columns of
 the second diagonalizing matrix in (\ref{diag}), we have
\be
\!\!\!W^\pm=\frac{1}{\sqrt{2}}({\cal{A}}^1 \mp i{\cal{A}}^2),\;\;Z=-\cos{\theta_W}{\cal{A}}^3
+\sin{\theta_W}{\cal{A}}^4,\;\;B=\sin{\theta_W}{\cal{A}}^3+\cos{\theta_W}{\cal{A}}^4, \ 
\label{bosons}
\ee
so (\ref{mk}) takes on the standard expression for the mass term as
\be
\kappa^2M_{IJ}{\cal{A}}^I_\a{\cal{A}}^J_\b\d^{\a\b}=\kappa^2\big(m^2_WW^+W^-+\frac{1}{2}m^2_ZZ^2\big)
 \ .
\label{massterm}
\ee

\no

\section{Discussion}

We may firstly re-focus on certain points in our preliminary overview in section 2.
The tilt (\ref{tilt}) of the frames, on occasion viewable as a `shake-up', 
is {\em not}  a perturbation
because it can be as steep or violent as it may, restricted only by the general requirement of at least $C^2$ 
differentiability. In addition to its obviously translational character,
the same tilt involves rotations too. The latter emerge as a consequence of the loss of the {\em global} 
hypersurface-forming property of the frames\footnote{ 
This can be better visualized in examples of empty rotating spacetimes in $d=4$, such as the Kerr solution
 or the G\"{o}del universe,
where the locally orthogonal-to-time $dt=0$ hypersurfaces do not mesh to allow globally consistent simultaneity.
This result is due to a tilt of the frames exactly as in (\ref{tilt}), with the spacetime vorticity proportional 
to the time derivative of that tilt. 
In the present context we have the gauge potentials ${\cal A^I}$ as tilt, with the field strengths ${\cal F^I}$ as
{\it vorticity}.} \cite{rs}, already hinted-to in relation to (\ref{m8}). 

The fundamental pair of torsion ${\cal T^A}$ and curvature ${\cal R^A_{\;\;B}}$, as introduced
by Cartan's structure equations in (\ref{torsion}) and (\ref{riemann}), is interrelated by the holonomy theorems to,
 respectively, 
the translations and rotations of the inhomogeneous group 
acting in the tangent spaces of whatever manifold they inhabit \cite{traut}, here the $M^8(\omega)$.
The effect of the specific translations and rotations induced by the tilt (\ref{tilt}) in their local action 
in $M^8(\omega)$ can be measured  (after the general case) with a construction process for  local geodesic quadrilaterals-to-be, by their failure to close; 
and with the parallel transport of a  tangent vector 
around a geodesic quadrilateral,  by its failure to return aligned to its initial direction.
The non-closure displacement, a {\em translation} element,  is a local measure of non-vanishing torsion ${\cal T^A}$.  
The non-alignment angle, a {\em rotation} element,  is a local measure of  non-vanishing  curvature  ${\cal R^A_{\;\;B}}$.
The first element accumulates linearly, as  two tangent vectors are being transported parallelly and under torsion-caused 
spiralling  from an initial vertex of the (failing-to-close) quadrilateral, 
along each-other's geodesics 
and gaining in `potential energy', pretty much as with a  winding coil. 
There is no differentiation involved in this transport so, in our case, the produced element has to be proportional to
 the ${\cal A^I}$ of the tilt which caused it.
The torsion is  therefore  likewise proportional, hence it contributes with the 
quadratic-in-${\cal A^I}$ mass term (`potential energy') to ${\cal L}_{HE}(\omega)$ 
in (\ref{lagm}).
The second element accumulates from the entire surface of a closed quadrilateral and it is one level of differentiation 
up, so, in our case, it expectedly involves $({\cal F})^2$, 
also viewable as a kinetic $({\it vorticity})^2$ term contribution to  ${\cal L}_{HE}(\omega)$. 
The interplay between translational and angular degrees of freedom 
 in the above
semi-qualitative interpretation follows in geometric elegance from their  ivolvement in the  rigorous 
 interrelation 
(by the mentioned  theorems on holonomy)  between
on one side the Noether currents and charges from (tangent-space) invariances of
the Poincar\`{e} group as {\em sources}  and, 
on the other,
the generated {\em fields} of torsion and curvature (\cite{traut}).

We may now argue that the  above {\mbox{\it co-existence}} of these two terms in ${\cal L}_{HE}(\omega)$, with
both having been created simultaneously and by common cause  (namely the tilt or shake-up (\ref{tilt}) of the frames, as seen) 
is {\em compulsory} in the sense that having one of these terms without the other
would mean direct violation of  the mentioned basic theorems on holonomy.  By the latter, this violation may also be viewed 
as that of  `mechanical-energy conservation', with  none of its kinetic and potential-energy contributions omittable.
In the underlying angular spring-oscillator  dynamics, the
anisotropic spring-constants of the vacuum are  identified with the eigenvalues of the mass matrix.
A consequence of this result is that nature must skip the middle step in the evolution scheme (\ref{evolv}), where 
the massless case (\ref{lagr}) 
cannot  be realized within the typical energy range of the standard EW interaction. Of course, at sufficiently high 
energies, as in the LHC experiment, 
phenomenology related to the existence of the middle step in (\ref{evolv}) may well be expected. 
In the conventional treatment  the massless case cannot be excluded from the outset. 
To elevate the  piecemeal addition of 
[\mbox{4(4-3)/2} graviton plus (8+7)  $SU(2)\times U(1)$  massless gauge-boson
and scalar]=17 independent states  in the symmetric Lagrangian (\ref{lagr})  to   
20 when the gauge bosons acquire mass {\em there}, the Goldstone's theorem is invoked for
the emergence of the 3 transverse states, accounted for by a simultaneous re-arrangement of the independent states 
of the Higgs fields
and the Goldstone bosons \cite{duff}. {\em Here}, in sharp contrast, the correct count of 20 follows directly as
\mbox{8(8-3)/2} for the $M^8(\omega)$ manifold while the massless case (\ref{lagr}) is disallowed at the mentioned energy
range, as seen. The eigenvalues in 
(\ref{diagm}) reproduce precisely  the spectrum of the
well-known experimental result, along with the parallel interpretation of the same mass spectrum as `spring-constants' of the vacuum, as explained.

Sufficiently close or above the very high energy scale of the massless EW interaction,
the ground-state manifold approach for the Kaluza-Klein gauge-theory vacuum (and any tilt of the frames therein) 
would have already been rendered inapplicable.
The so-called {\em cylinder condition} would have to be abandoned for a more primitive state of the $M^8(\omega)$ 
manifold.  
Under such considerations, and if we
were to restrict ourselves to the gravitational and EW interactions, it is tempting to conjecture or speculate 
that the  $S^3\times S^1$ sector of the topology in 
(\ref{m8}) might be expected to emerge with even deeper importance, due its {\em mixmaster} behavior \cite{rs}. 
This refers to Misner's  profoundly non-linear dynamics which could turn {\em turbulent}\footnote
{We choose the term `turbulent' to mean  chaotic and micro-causality violating vacuum dynamics, hence one 
with classical variables which would violate Bell's inequalities, just like 
quantum mechanical ones. The latter character should also prescribe a specially needed handling of causal
transforms, e.g., Fourier expansions, which would in principle be meaningless in such environments.} 
in the presence of  sufficiently steep  potential walls. 
Then, through Kasner-like bounces on them, at Planck-scale frequencies, such dynamics may enforce  isotropy and homogeneity on the geometry actually or, better, effectively.
The latter case may be realized as seen from a
sufficiently longer time scale or as averaged at a much-lower energy scale. 
It may then be possible that what we assume as the cylinder condition, or employ as a static 
$S^3\times S^1$ sector in $M^8(0)$ at the much lower EW energy scale of ${\cal L}_{HE}(\omega)$ in (\ref{lagm}),
 is the {\em effectively} static presence 
of an actually turbulent vacuum dynamics
at the much deeper Planck scale. Such studies could justify the cylinder condition  and illuminate the question 
of the classical stability of the ground-state manifold, possibly in relation to its quantum mechanical phenomenology.

If encouraged by the present development (and before any dimensional enlargement of the geometry), 
one might investigate other types of components of the torsion tensor, in the context of holonomy as mentioned, and in relation 
to the  invariants (spin and mass) and representations of the  Poincar\`{e} group.  
In the wider area of the present work there have been earlier contributions with torsion-related aspects and
potentially observable effects and testable consequences \cite{bat}.
Those which may be carried over to the present context  include certain Aharonov-Bohm type of gravito-EW
 interferences and gyro-magnetic effects, 
as well as a lower bound for the violation of the principle of equivalence,
expectedly at roughly the order of 
$L_1^{-1}/L_o^{-1}\sim 1:10^{17}$
(existing tests are negative down to $\sim 1:10^{12}$). Additional effects may be revealed as, e.g.,
 with the study of couplings and geodesics
independently or under (\ref{ttcom}), now understood with $\theta=\theta_W$.

\vskip .03in
\no
I am grateful to A.~A.~Kehagias for  discussions. And to the ever-inspiring contributors to our {\em milieux}
for ideas, prompts and ways to go after the awesome (frightening, too) taste of pure marble-stone soup!

\end{document}